\documentclass[a4paper,fleqn,5p,times]{cas-dc}

\usepackage{amssymb}
\usepackage{amsmath}

\usepackage{algorithmic}
\usepackage{graphicx}
\usepackage{textcomp}
\usepackage{xcolor}

\usepackage{xurl}

\usepackage{booktabs}
\usepackage{multirow}
\usepackage{hyperref}
\usepackage{enumitem}
\def\BibTeX{{\rm B\kern-.05em{\sc i\kern-.025em b}\kern-.08em
    T\kern-.1667em\lower.7ex\hbox{E}\kern-.125emX}}

\usepackage[numbers]{natbib}

\usepackage[acronym]{glossaries}
\makeglossaries

\newacronym{ai}{AI}{Artificial Intelligence}
\newacronym{ml}{ML}{Machine Learning}
\newacronym{siem}{SIEM}{Security Information and Event Management}
\newacronym{soc}{SOC}{Security Operation Center}
\newacronym{its}{ITS}{Intrusion Tolerant System}
\newacronym{soar}{SOAR}{Security Orchestration, Automation, and Response}
\newacronym{misp}{MISP}{Malware Information Sharing Platform}
\newacronym{apt}{APT}{Advanced Persistent Threat}
\newacronym{cti}{CTI}{Cyber Threat Intelligence}
\newacronym{cis}{CIS}{Cybersecurity Information Sharing}
\newacronym{fl}{FL}{Federated Learning}
\newacronym{mhe}{MHE}{Multiparty Homomorphic Encryption}
\newacronym{dp}{DP}{Differential Privacy}
\newacronym{smpc}{SMPC}{Secure Multiparty Computation}
\newacronym{zkp}{ZKP}{Zero-Knowledge Proofs}
\newacronym{pgi}{PGI}{Private Graph Intersection}
\newacronym{psi}{PSI}{Private Set Intersection}
\newacronym{ciso}{CISO}{Chief Information Security Officer}
\newacronym{ioc}{IOC}{Indicators Of Compromise}
\newacronym{osint}{OSINT}{Open-Source Intelligence}
\newacronym{hids}{HIDS}{Host-based Intrusion Detection System}
\newacronym{vm}{VM}{Virtual Machine}
\newacronym{legion}{LegionITS}{Linked Entities for Governance of Intrusion and Operational Norms of Intrusion Tolerant Systems}
\newacronym{ids}{IDS}{Intrusion Detection System}
\newacronym{nist}{NIST}{National Institute of Standards and Technology}
\newacronym{he}{HE}{Homomorphic Encryption}
\newacronym{os}{OS}{Operating System}

\begin{document}
\shorttitle{\gls{legion}}
\shortauthors{T. Freitas et~al.}



\author[1]{Tadeu Freitas}[orcid=0000-0001-5573-2434]
\ead{tadeufreitas@fc.up.pt}\cormark[1]
\author[1]{Carlos Novo}[orcid=0009-0003-0094-5565]
\ead{carlos.novo@fc.up.pt}
\author[1,2]{Manuel E. Correia}[orcid=T0000-0002-2348-8075]
\ead{mdcorrei@fc.up.pt}
\author[1,3]{Rolando Martins}[orcid=0000-0002-1838-1417]
\ead{rmartins@fc.up.pt}

\cortext[1]{Corresponding author}

\affiliation[1]{
    organization={Departamento de Ciências de Computadores, Faculdade de Ci\^{e}ncias, Universidade do Porto},
    postcode={Porto}, 
    country={Portugal}
}
\affiliation[2]{
    organization={CRACS - INESC TEC},
    postcode={Porto}, 
    country={Portugal}
}

\affiliation[3]{
    organization={Safehelm},
    postcode={Porto}, 
    country={Portugal}
}

\title [mode = title]{\acrshort{legion}: A Federated Intrusion-Tolerant System Architecture
}




\begin{abstract}
The growing sophistication, frequency, and diversity of cyberattacks increasingly exceed the capacity of individual entities to fully understand and counter them.
While existing solutions, such as \gls{siem} systems, \gls{soar} platforms, and \gls{soc}, play a vital role in mitigating known threats, they often struggle to effectively address emerging and unforeseen attacks.
To increase the effectiveness of cyber defense, it is essential to foster greater information sharing between entities; however, this requires addressing the challenge of exchanging sensitive data without compromising confidentiality or operational security.

To address the challenges of secure and confidential \gls{cti} sharing, we propose a novel architecture that federates \glspl{its} and leverages concepts from \gls{misp} to empower \glspl{soc}. 
This framework enables controlled collaboration and data privacy while enhancing collective defenses. 
As a proof of concept, we evaluate one module by applying \gls{dp} to \gls{fl}, observing a manageable accuracy drop from $98.42\%$ to $85.98\%$ (average loss $12.44\%$) while maintaining reliable detection of compromised messages.
These results highlight the viability of secure data sharing and establishes a foundation for the future full-scale implementation of \acrshort{legion}.

\end{abstract}









\maketitle

\glsresetall

\section{Introduction}\label{sec:introduction}
In an increasingly connected and digitally dependent world, ensuring critical systems' confidentiality, integrity, and availability is paramount~\cite{cawthra2020data}. 
\gls{its} enhance cybersecurity by emphasizing resilience over prevention, ensuring systems can continue functioning correctly (with limited capability) despite successful attacks or compromises against known intrusions and vulnerabilities~\cite{verissimo2003intrusion, wang2003sitar}.

These systems are typically decentralized and incorporate replication, detection, and recovery mechanisms to maintain resilience and fault tolerance.
However, it is possible to compromise them with more sophisticated adversaries who continuously adapt and evolve to bypass defenses through zero-day vulnerabilities/attacks~\cite{bleepingcomputer2025}.

Recent attacks on the blockchain~\cite{zhang2023fairness} and the cloud~\cite{xsa396} have exposed the fragility of currently deployed frameworks and solutions in the face of malicious adversaries. 
This development underscores the potential for integrating intrusion-tolerant systems (ITS) into cybersecurity frameworks. Despite the growing interest in ITS, ongoing efforts aim to translate theoretical approaches~\cite{verissimo2003intrusion, verissimo2006intrusion, sousa2009highly, moniz2008ritas, correia2013bft} into practical, real-world solutions~\cite{garcia2019lazarus, freitas2023skynet, hammar2024intrusion} to mitigate the impact of such attacks.

However, a standalone \gls{its} may not be sufficient to address sophisticated and complex threats. 
Specifically, when considering attacks that can undermine the safety invariants and avoid detection through Cross-platform Zero-days~\cite{lindorfer2013poster}, affect node rejuvenation and recovery~\cite{sayed2022cyber}, anomaly evasion~\cite{cheng2011evasion}, and log forgery/tampering~\cite{liu2023data}.

Adopting a ``combine and conquer'' approach offers a more tailored solution by integrating existing knowledge with insights acquired by other entities, i.e., combining the knowledge gathered by \glspl{its}.
This is critical to mitigate these threats~\cite{johnson2017cyber}.
In an ever-changing threat landscape~\cite{dave2023new}, retrieving and sharing information on emerging threats and novel defenses can enable systems to adapt more effectively. 
However, sharing sensitive data or control mechanisms in environments with multiple organizations using \glspl{its} introduces confidentiality and trust concerns.
To address these challenges, we hypothesize that the Federation paradigm could enable independent entities to collaborate against threats by exchanging insights, decisions, or aggregated data about malicious actors, while preserving their autonomy and maintaining data privacy. 

Federated \gls{its} aims to enhance scalability, promote secure collaboration while maintaining data autonomy, privacy, and confidentiality, and improve the system's resilience against \gls{apt}.

The objective of this work is to address the following research questions:
\begin{enumerate}[label=Q\arabic*:]
    \item How can the Federation paradigm be integrated into \glspl{its}, and what challenges might arise?
    \item How can information sharing be effectively managed at both inter-level (between entities) and intra-level (within an entity)?
    \item What types of information can be safely and effectively shared between organizations without compromising privacy or security?
    \item In what ways does newly acquired information benefit an individual entity within a collaborative system?
    \item How can shared information improve the overall resilience of a system or interconnected systems?
\end{enumerate}

This paper aims to provide the following contributions:
\begin{itemize}
    \item Propose \gls{legion} a novel architecture that integrates the Federation paradigm with \glspl{its}, enabling entities with separate clusters to share knowledge internally.
    \item Develop mechanisms to enable coordination between \glspl{legion}, fostering organization collaboration for exchanging knowledge acquired within their respective domains.
    This will involve introducing a dedicated distributed ledger-based solution to securely store and manage the collected knowledge, ensuring it is a reliable resource for existing and new participants.
    \item Integrate the acquired knowledge with public malware information and threat-sharing platforms, enhancing the global cybersecurity ecosystem.
\end{itemize}

The rest of the paper is organized as follows. 
Section~\ref{sec:background} offers essential background information, enabling readers to understand the terminology and concepts referenced throughout the article.
Section~\ref{sec:related_work} addresses the state of the art regarding research on \gls{cti}, \gls{cis}, and \glspl{its}.
Section~\ref{sec:architecture} details the architecture of \gls{legion}, introduces a theoretical-practical scenario that illustrates its benefits in threat mitigation, and provides a practical scenario to demonstrate the trade-offs of employing \gls{fl} with \gls{dp}.
Section~\ref{sec:evaluation} presents an evaluation of the architecture's response to expected scenarios, using the Scenario-Based Architecture Analysis Method (SAAM)\cite{1019479, kazman2002scenario}, which explores how the architecture meets operational requirements and adapts to potential future changes. 
Section~\ref{sec:discussion} discusses the research questions and connects them to the proposed approach and architectural decisions.
Finally, Section~\ref{sec:conclusion} provides a conclusion regarding the research done and a perspective for future work.

\section{Background}\label{sec:background}
Before presenting the related work and \gls{legion}, a comprehensive background is provided that covers the core concepts of \gls{its}, \gls{cis}, \gls{cti}, and the underlying principles of federated architectures.\\

\noindent\textbf{Intrusion Tolerant System~\cite{verissimo2003intrusion}} is a fault-tolerant architecture designed to maintain both operational functionality and security even when intrusions occur.
These systems prevent intrusions from causing total system failure or exposing sensitive data. 
This resilience is achieved through mechanisms that enforce confidentiality, integrity, and availability. 
In practice, \glspl{its} implement strategies such as redundancy, diversity, active self-healing, and dynamic reconfiguration.
Consequently, even if an attacker compromises part of the system, the overall system remains secure and operational, effectively containing and mitigating the intrusion.\\

\noindent\textbf{Federated Architectures~\cite{heimbigner1985federated}} are decentralized computing models that enable autonomous systems, organizations, applications, or components to collaborate while maintaining their independence. 
This approach balances local autonomy with global coordination, facilitating efficient data sharing, scalability, and interoperability across diverse systems~\cite{teixeira2025understanding}.

Key features include decentralization, where each unit operates independently without centralized control; interoperability, allowing seamless interaction between systems with different technologies or standards; scalability, enabling growth through the integration of new systems without disrupting existing ones; and standardization, which ensures consistent communication and secure data exchange through shared protocols.
Federated architectures are ideal for managing complex systems with diverse stakeholders while preserving flexibility and autonomy.\\

\noindent\textbf{Cyber Threat Intelligence~\cite{wagner2019cyber}} is collecting, analyzing, and disseminating information about cyber threats to help stakeholders understand potential risks, threat actors, attack methods, and vulnerabilities.
This transforms raw data into actionable insights, allowing security teams to take defensive measures against current and emerging cyber threats proactively.

\gls{cti} typically includes details on threat actors' tactics, techniques, and procedures, \gls{ioc}, and attack campaigns.
This can be categorized into strategic, tactical, operational, and technical.
\gls{cti} are necessary to improve cybersecurity defenses, prioritize resources, mitigate risks, and enable informed decision-making across organizations.\\

\noindent\textbf{Cybersecurity Information Sharing~\cite{pala2019information}} is a method to exchange data, insights, and intelligence regarding cyber threats, vulnerabilities, incidents, and defensive measures between organizations, sectors, or governments.
The objective is to collectively improve cybersecurity resilience and allow participants to identify, mitigate, and respond to cyber threats.

It considers automated sharing systems, collaborative frameworks, and policies that guarantee shared information's confidentiality, integrity, and utility while providing privacy and regulatory requirements.
This is used to prevent widespread cyberattacks across diverse entities.\\


\noindent\textbf{Synergy and trade-offs}

The integration of \gls{its}, \gls{cti}, \gls{cis}, and federated architectures can be combined to increase cybersecurity resilience across organizations. 
This synergy is achieved by establishing a decentralized framework that facilitates collaborative threat mitigation, adaptive infrastructure, and shared intelligence.

Federated architectures serve as the structural backbone of this integration, enabling \gls{its} systems to operate collaboratively without reliance on centralized entities.
This decentralized approach allows organizations to exchange threat intelligence and coordinate defensive strategies while maintaining autonomy.
Within this framework, \gls{its} usually provides raw data, such as system logs, which are subsequently processed by \gls{cti} systems into actionable insights. 
These insights detail adversary tactics and techniques, empowering organizations to adopt proactive defense measures.

In addition to sharing \gls{cti}, the secure exchange of \gls{cis} offers organizations valuable information beyond threat indicators. 
This includes details on security controls, configuration best practices, incident response procedures, and overall cybersecurity posture. 
Sharing this broader range of information deepens collective understanding of the threat landscape and supports better decisions regarding resource allocation and infrastructure improvements.

Despite these advantages, several trade-offs exist in implementing federated systems. 
Data standardization remains a critical challenge, as federated systems require interoperable formats to effectively correlate threats across diverse datasets. 
Trust-building among participating entities is another essential factor for success; mechanisms like accountability modules, exemplified by Melicertes~\cite{IRIS2021}, can foster transparency and trust within federated frameworks.

Operational complexity also poses challenges due to the high volume of data generated by \gls{its} and \gls{cti} feeds, which can overwhelm analytical capacities.
Furthermore, regulatory risks emerge when sharing information across jurisdictions, particularly when data sovereignty laws and privacy issues conflict with collaborative defense efforts.
Addressing these challenges requires careful consideration of governance frameworks and technological solutions to balance collaboration with compliance.

\section{Related Work}\label{sec:related_work}

Several research efforts have explored federated architectures, \gls{its}, \gls{cti}, and \gls{cis} individually; however, to the best of our knowledge, there is a lack of work that conceptually integrates these domains within a unified high-level architecture.
Nevertheless, work across these areas offers valuable design perspectives. 
This section highlights key contributions that have influenced the architecture design of \gls{legion}.

In the space of privacy-preserving \gls{cti} sharing, Freudiger et al.~\cite{freudiger2014privacy} introduced the SIC framework, combining utility-based decision-making with \gls{psi} to enable selective and secure collaboration.
Van de Kamp et al.~\cite{van2016private} proposed a cryptographic approach that allows \glspl{ioc} and intrusion sightings to be shared without revealing sensitive details to a central aggregator. 
DNSBloom, developed by Van Rijswijk-Deij et al.~\cite{van2019privacy}, uses Bloom filters~\cite{tarkoma2011theory} to support malicious DNS query detection while minimizing the risk of individual data exposure.

Badsha et al.~\cite{badsha2019privacy} designed a collaborative framework using \gls{he} to build a global decision tree from encrypted local data.
While the approach preserves confidentiality, it relies on a central server, a single point of failure that \gls{legion} explicitly avoids.

A more distributed approach was presented by Trocoso-Pastoriza et al.~\cite{trocoso2022orchestrating}, who introduced a federated \gls{cti}-sharing platform that collects threat data from distributed sources, e.g., vulnerabilities, incidents, \glspl{ioc}.
It combines \gls{mhe}~\cite{mouchet2023multiparty}, \gls{fl}, and \gls{dp}, enabling the training of threat prediction models without compromising raw data.

Preuveneers et al.~\cite{preuveneers2023privacy} proposed a flexible framework for \gls{cti} producers and consumers using polyglot persistence and analysis. 
It constructs correlation graphs from diverse threat intelligence sources, such as \glspl{ioc}, vulnerabilities, and adversary profiles, while applying anonymization, hashing, and encryption to preserve privacy.
Policy-driven data sharing ensures fine-grained access control across organizations.

In the context of \gls{cis}, Fisk et al.~\cite{fisk2015privacy} introduced three foundational principles for privacy-preserving cybersecurity information sharing: Least Disclosure, to minimize unnecessary data exposure; Qualitative Evaluation, to incorporate legal and ethical considerations alongside technical metrics; and Forward Progress, to encourage collaborative, constructive data sharing practices.

Vakilinia et al.~\cite{vakilinia2017privacy} proposed a framework that uses blind signatures, group signatures, and \gls{zkp} to protect participant anonymity while supporting accountability.
While it safeguards identities, the threat data itself remains unencrypted.
The system includes dispute resolution and incentive mechanisms to prevent free-riding and promote trustworthy participation.

Fuentes et al.~\cite{de2017pracis} introduced PRACIS, a privacy-aware analytics platform built on the STIX~\cite{stix21} standard. 
It combines \gls{he} and format-preserving encryption\cite{bellare2009format} to prevent linkability between threat reports and their sources, even under server compromise. 
The system follows a publisher-subscriber model and uses HMACs for integrity checks, focusing primarily on efficient summary statistic generation.

CYBEX-P, developed by Sadique et al.~\cite{sadique2019}, takes a broader view by combining trusted computing, public key cryptography, and blind processing to anonymize both data and users. 
Encryption mechanisms differ based on whether the data is in cache or archive, and blind processing removes any identifying details before distribution.

Research on federated architectures has explored a variety of concepts across different domains. 
Heimbigner and McLeod~\cite{heimbigner1985federated} introduced one of the earliest frameworks for integrating heterogeneous and autonomous database systems. 
Their architecture enables dynamic participation by employing a shared schema and a federated dictionary, which together facilitate interoperability without relying on centralized control.

Wijegunaratne and Fernandez~\cite{wijegunaratne2000federated} expanded this concept in enterprise settings, introducing an event-driven, domain-aligned architecture built around three core patterns: Federation, for grouping applications into logical domains; Dependency Separation for managing cross-domain processes; and Interface Connection, for defining inter-domain communication. 
Their earlier work~\cite{wijegunaratne2012distributed} established the idea that federated systems must support structured, controlled data exchange without disrupting the autonomy of their members.

Rizk et al.~\cite{rizk2021graph} introduced a graph-based \gls{fl} to overcome the limitations of a centralized federated learning model.
Their decentralized topology connects multiple servers and clients, improving scalability and resilience to failure. 
Privacy is maintained through \gls{dp} and cryptographic masking, modeled as additive noise. 
Under convexity and Lipschitz conditions, they show that learning performance remains close to that of non-private baselines.

In the \gls{its} domain, Veríssimo et al.~\cite{verissimo2003intrusion} argued for a shift away from purely preventative security measures toward architectures that assume compromise and prioritize containment. 
Their model treats resilience as equally important as security, enabling systems to continue operating even under attack.

Follow-up work in \gls{its}~\cite{bangalore2009securing, garcia2019lazarus, freitas2023skynet, hammar2024intrusion} builds on this by making systems more adaptive and proactive.
These newer designs support dynamic node recovery, architectural reconfiguration, and automated response capabilities. 
Many integrate external intelligence sources, including \gls{osint} feeds and social media data, into protected control planes for continuous learning.

Modern \glspl{its} not only consumes \gls{cti} but also produces it. 
Insights into attacker behavior, mitigation strategies, and threat patterns are stored and fed into control planes capable of automation.
This reduces the need for human intervention and significantly shortens the threat exposure window.

Collectively, these components outline the conceptual basis for \gls{legion}: a federated \gls{its} architecture intended to promote privacy-preserving threat intelligence sharing and support more coordinated and automated threat mitigation across distributed organizations.
This approach emphasizes autonomy, resilience, and scalability by design, and highlights key architectural principles and mechanisms that could inform future developments in this area.

\section{System and threat model}

This section introduces the key stakeholders in the federation and outlines the threat model considered during the architectural design.
It then presents the system model and its underlying assumptions.

\subsection{Stakeholders and Threat Model}

In \gls{legion}, each organizational stakeholder is assumed to perform multiple roles within the federated infrastructure. 
These roles encompass data provision, data processing, and data consumption.

As data providers, stakeholders include intelligence agencies, cyber defense units, and internal systems such as deployed \gls{misp} instances or \gls{osint} aggregators that collect and generate \gls{cti}. 
Any data shared within the federation is subject to strict privacy and confidentiality requirements.
To safeguard sensitive information, data providers are expected to enforce protective measures-access control policies, data minimization strategies, anonymization techniques, encryption schemes, and policy-driven sharing agreements. 
Despite these protections, providers may vary in their behavior. 
Some are assumed to be honest, following the defined protocols.
Others may be semi-honest, i.e., they follow the protocol but attempt to infer sensitive information from shared data.
A compromised stakeholder may behave maliciously, submitting falsified, manipulated, or privacy-violating intelligence to degrade the system’s reliability or undermine trust in shared intelligence.

Stakeholders also function as data processors, responsible for the transformation, correlation, and enrichment of \gls{cti}. 
This may involve analyzing internal telemetry, participating in distributed learning workflows, or contributing to federated inference tasks.
Processors may behave honestly, semi-honestly, or maliciously, with the latter potentially tampering with data flows, injecting adversarial examples, or violating processing logic for inference or disruption.

In the role of data consumers, stakeholders query intelligence repositories, either for operational monitoring or automated response. 
These consumers include human analysts, detection platforms, and response orchestration frameworks that act upon the retrieved data to trigger alerts, apply mitigations, or support defensive decision-making. 
Data consumers are assumed to behave either honestly or semi-honestly, without the ability to inject or modify upstream data but possibly attempting to infer insights from aggregated intelligence.

The communication model assumes that interactions across the federation take place over a partially synchronous and unreliable network. 
Messages exchanged between participants may experience delays, losses, duplication, or reordering. 
This model accounts for both benign network faults and potential adversarial interference at the communication layer.

Under the assumption of a win-win collaboration among $N$ networked organizations, \gls{legion} assumes that each participant simultaneously acts as a data provider and data consumer. 
Each organization possesses sufficient computational capacity—whether in-house or outsourced—to support both local analysis and participation in collaborative processing over the federated intelligence space. 
These capabilities enable a decentralized, resilient, and privacy-preserving environment for cross-organizational cyber defense.

\subsection{System Model}

The \gls{legion} framework models a federated system composed of $N$ autonomous organizations, each operating one or more local \glspl{its} within their internal environments. 
These systems may be deployed at different levels of granularity, such as per department, team, or operational unit, enabling internal segregation and compartmentalization.
This local deployment structure allows organizations to build intra-organizational federations that reflect internal trust boundaries and policy domains.

Each local \gls{its} is responsible for collecting, processing, and acting upon \gls{cti} relevant to its specific domain.
Local \gls{its} instances within an organization can collaborate under policy-controlled conditions, enabling shared analysis and mutual reinforcement while preserving data sovereignty and isolation across varying confidentiality classifications.

This model is extended by enabling federated collaboration across organizational boundaries.
The framework supports the secure sharing of \gls{cti}, joint execution of distributed analyses, and sharing resilience strategies between organizations.
These interactions are governed by policies that define what information can be shared, with whom, and under which conditions, ensuring compliance with privacy, legal, and strategic requirements.

The system supports federated operations such as cross-domain correlation, collaborative anomaly detection, and privacy-preserving aggregation.
These are assumed to be performed through distributed protocols that do not employ centralized coordination and are resilient to network faults and partial participation.

Communication between \gls{its} instances—whether intra- or inter-organizational—occurs over an unreliable, partially synchronous network where messages may be delayed, lost, duplicated, or delivered out of order.
The system is designed to operate effectively under such conditions and assumes no inherent trust in the communication layer.

\gls{its} deployments within each organization are expected to integrate with existing infrastructure such as \gls{cti} platforms, e.g., \gls{misp}, logging systems, detection engines, and response tools. 
Federation mechanisms rely on standardized data representations, e.g., STIX, and exchange protocols, e.g., TAXII~\cite{taxii21}, ensuring interoperability across heterogeneous environments.

The system model reflects the behavioral assumptions introduced in the threat model, allowing for a mix of cooperative and adversarial participants while maintaining operational correctness, confidentiality, and resilience.

\begin{figure*}[htbp]
    \centering
    \includegraphics[width=\linewidth]{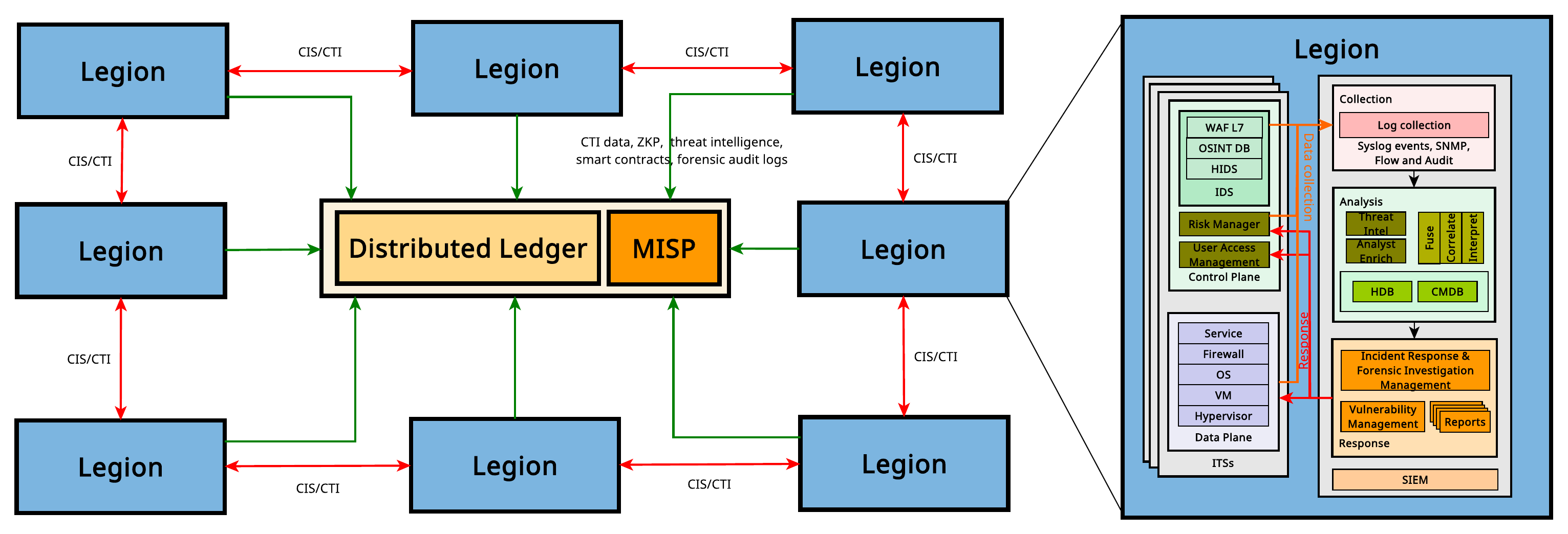}
    \caption{\gls{legion}’s architecture, illustrating both internal and inter-organization sharing.}
    \label{fig:arch}
\end{figure*}

\section{Architecture}\label{sec:architecture}

This section introduces \gls{legion}'s architecture, providing an overview of its core components and design principles.
We demonstrate its practical application with a case study on how the architecture handles a Zero-Day Vulnerability scenario.
Lastly, we present an evaluation comparing standard \gls{fl} and \gls{fl} enhanced by \gls{dp}, analyzing the trade-offs between system performance and privacy protection.\\

\gls{legion} is a federated threat intelligence system, depicted in Figure~\ref{fig:arch}, that utilizes a two-level federation to optimize both internal and external sharing of threat intelligence.
The system’s design fosters collaboration while ensuring open access to critical cybersecurity insights.

To clarify the system’s core components, we begin by discussing how \gls{legion} integrates with modern \glspl{its}.
The architecture emphasizes the relationship between \gls{cti}-producing and \gls{cti}-consuming elements, which support intelligence-driven cybersecurity operations.

Advancements in \gls{its} and \gls{siem} systems have influenced the integration strategy, focusing on separating control plane and data plane functions.
Within the control plane, the \gls{ids} includes a Web Application Firewall (WAF) operating at Layer 7. 
WAF provides advanced features such as deep malware inspection, binary analysis, and the attribution of attacks to known threat actors, including \glspl{apt}.

The \gls{ids} includes an \gls{osint} module that collects threat data from public repositories for quick integration into analysis workflows.
Complementary to this, the \gls{hids} collects detailed telemetry on system-level intrusions, improving the accuracy of local threat detection.

A dedicated risk management component continuously evaluates the \gls{its}'s exposure to cyber threats by synthesizing intelligence on vulnerabilities, exploits, and system configurations.
This assessment guides recommendations for secure and resilient configurations, which can be shared both internally and externally to foster collaborative defense.

User Access Management (UAM) plays a crucial role in detecting behavioral anomalies and potential privilege misuse. 
By continuously monitoring authentication logs, access patterns, and escalation events, UAM systems can identify technical indicators, such as anomalous IP addresses, unusual login times, geolocation mismatches, and repeated failed login attempts, that often precede or accompany malicious activity.
These indicators align with known adversarial techniques, including privilege escalation and lateral movement, as described in frameworks like MITRE ATT\&CK~\cite{strom2018mitre}.
Furthermore, analyzing access patterns at the user or departmental level helps pinpoint which parts of an organization are being targeted, supporting both operational threat responses and strategic risk assessments. 
While the effectiveness of this approach depends on implementation, it is widely adopted in cybersecurity operations due to its proven value in early threat detection.

Telemetry is collected from a variety of system components across the data plane, shown in Figure~\ref{fig:arch}, which includes hypervisors, virtual machines, operating systems, firewalls, and application services.
This data supports post-event forensic analysis and the development of behavioral baselines for anomaly detection.

Combination with \gls{siem} systems allows for the transformation of raw telemetry into actionable threat intelligence. 
The \gls{siem} pipeline aggregates data from distributed \gls{its} while enforcing robust privacy and confidentiality protections. 
Sensitive fields are masked or pseudonymized, logs are encrypted during transmission and storage, and data collection is secured through protocols such as SNMPv3. 
Role-based access controls regulate data access, and the data is sanitized to anonymize network flow endpoints and remove host-specific details.
Retention policies are implemented to manage the lifecycle of collected data, ensuring that analytic processes uphold individual or organizational privacy, support compliance with regulations such as GDPR, and accommodate varying data retention periods as required by policy or law.

Once anonymized, the data is enriched and interpreted by analysts. 
Through correlation and contextual analysis, threat intelligence is generated and validated.
This process is supported by a Hardware Database (HDB), which stores structured information about system assets. 
The HDB facilitates network discovery, inventory management, baseline security assessments, and the integration of Configuration Management Databases (CMDBs). 
CMDBs track detailed records of IT assets and their interdependencies, aiding in impact analysis, change tracking, incident correlation, and audit reporting. 
Together, these systems enable informed prioritization of vulnerabilities and effective scoping of security incidents.
With this contextual understanding of threats, the system can generate and deploy targeted mitigation strategies.

The resulting intelligence drives targeted responses across subsystems, with risk configurations updated via the risk manager and identity roles or access privileges adjusted based on findings from user access monitoring.
Intelligence is also shared with trusted peers and community partners to enhance collective situational awareness and improve cyber resilience.

\gls{legion} is designed to enable secure and efficient threat intelligence exchange across participating entities, supporting both intra- and inter-organizational collaboration. 
Given the sensitive nature of the data, the system enforces stringent privacy and confidentiality guarantees, providing a reliable and verifiable intelligence source for \glspl{ciso} and security teams while facilitating integration into automated threat detection and response pipelines.

To achieve these objectives, \gls{legion} integrates advanced cryptographic and privacy-preserving techniques, including \gls{mhe}, \gls{dp}, \gls{zkp}, Encrypted Anonymized Threat Intelligence, and protected audit logs with forensic evidence.
These techniques ensure secure exchange of threat intelligence while preserving privacy, fostering a robust framework for federated cybersecurity collaboration.

\begin{figure}[htp]
    \centering
    \includegraphics[width=\columnwidth]{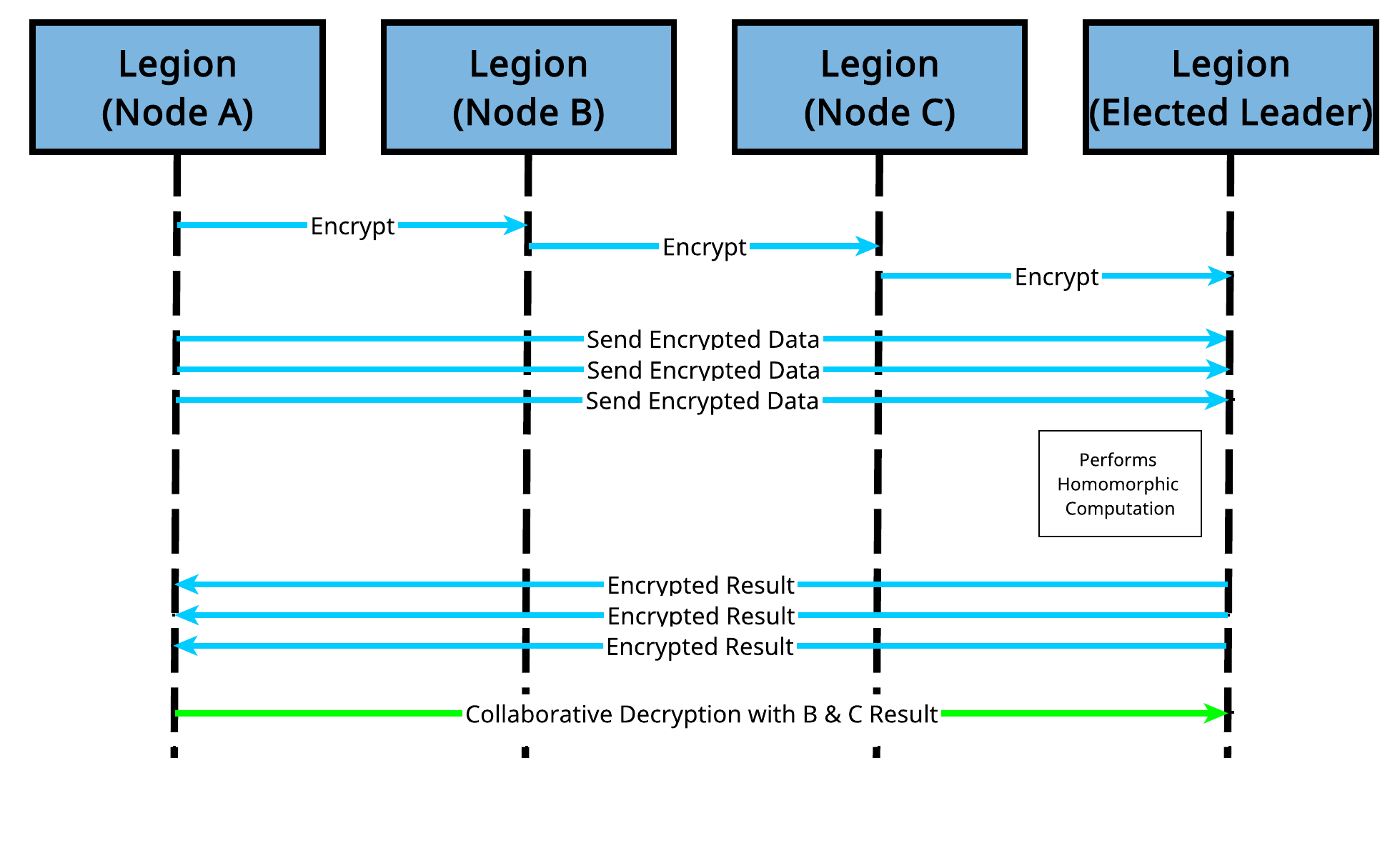}
    \vspace{-1cm}
    \caption{}
    \label{fig:mhe}
\end{figure}

\gls{mhe} allows computations on encrypted data without decryption, ensuring that model training and analysis can be conducted while preserving confidentiality, presented in  Figure~\ref{fig:mhe}.
\gls{dp} introduces controlled statistical noise to protect individual data points from inference attacks.
This approach mitigates data poisoning risks while preserving the utility of the shared intelligence.

\begin{figure}[htp]
    \centering
    \includegraphics[width=\columnwidth]{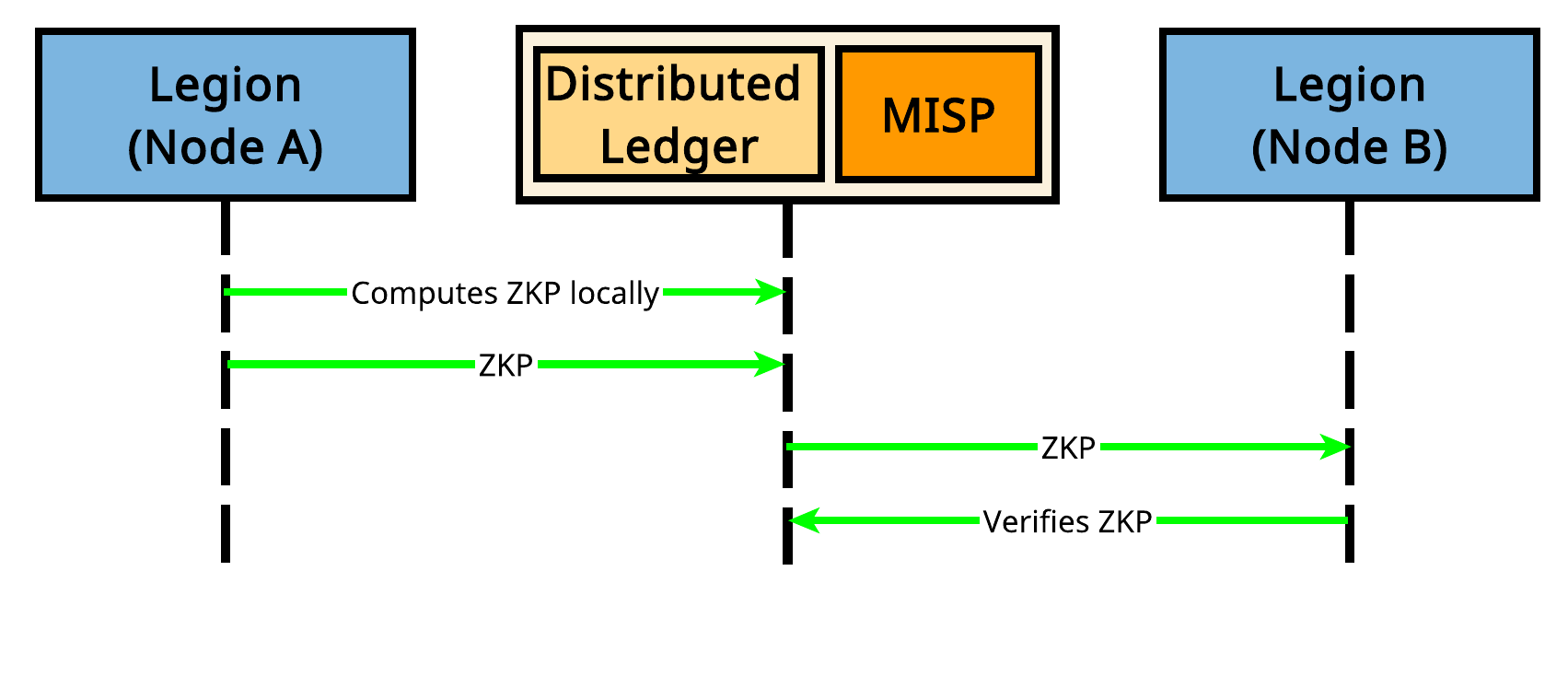}
    \vspace{-1cm}
    \caption{Dataflow for \gls{zkp} for Data Sharing.}
    \label{fig:zkp}
\end{figure}

\gls{zkp} enables organizations to verify exposure without disclosing sensitive information, such as compliance with mitigation policies or vulnerability exposure.
This is achieved through a challenge-response mechanism, demonstrated in Figure~\ref{fig:zkp}, which verifies the validity of security claims while ensuring that underlying data remains confidential.

Encrypted Anonymized Threat Intelligence ensures secure exchange of critical cybersecurity information, including \glspl{ioc} such as MITRE ATT\&CK techniques, blacklisted IP addresses, and malware signatures, while preserving data source anonymity.
Protected audit logs and forensic evidence guarantee the integrity and traceability of security events, facilitating forensic analysis without compromising sensitive data.

Protected audit logs and forensic evidence are essential in identifying patterns indicative of security breaches, thereby strengthening threat detection capabilities. 
By serving as an additional data source for \gls{fl} models and enabling hash-based integrity verification, these logs enhance the system's accuracy and speed in anomaly detection.

\begin{figure}[htp]
    \centering
    \includegraphics[width=\columnwidth]{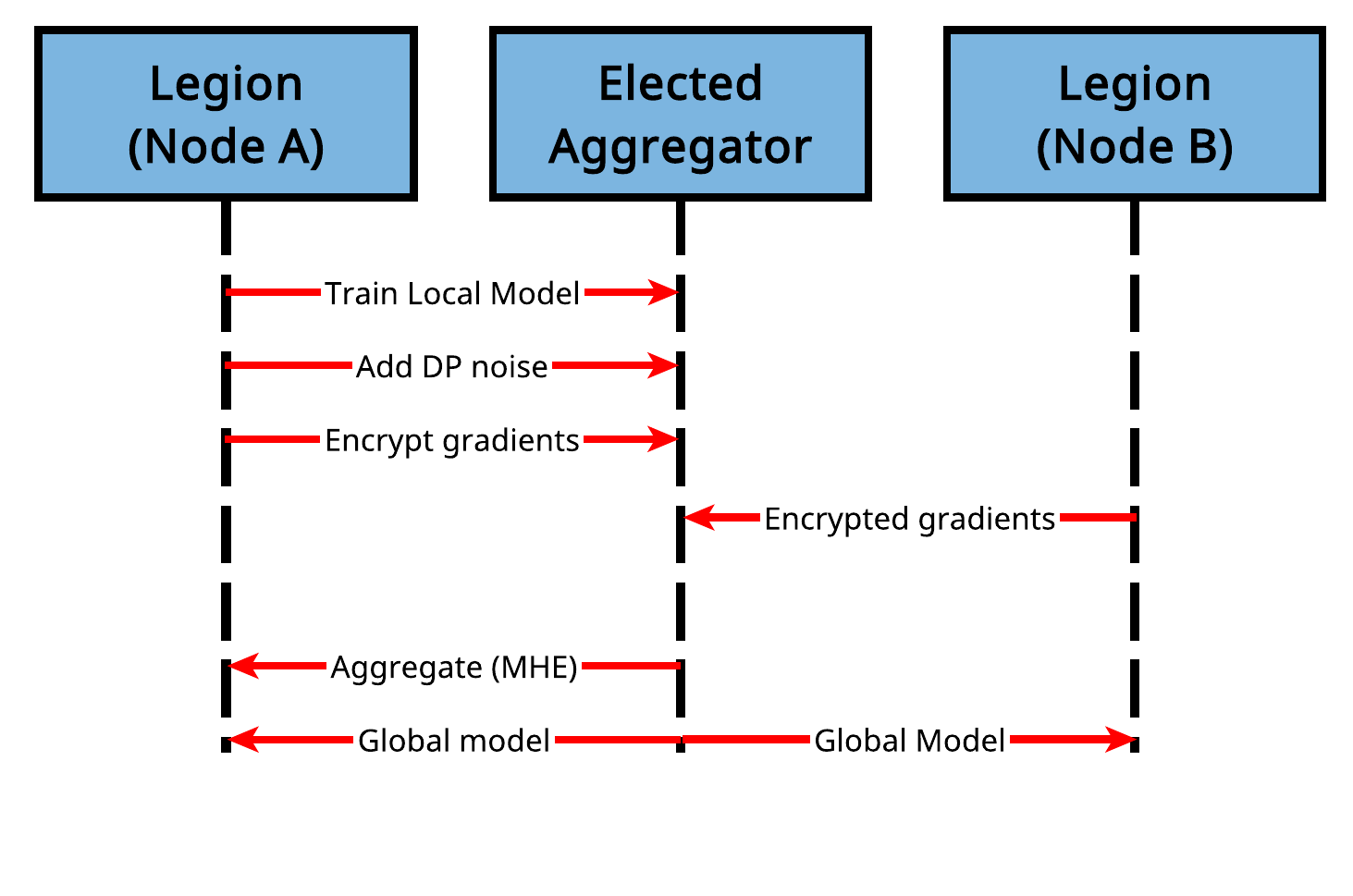}
    \vspace{-1cm}
    \caption{Dataflow for \gls{fl} combined with \gls{dp}, and further secured by applying \gls{mhe}.}
    \label{fig:mhe-dp-fl}
\end{figure}

To further enhance its analytical capabilities, the architecture integrates two distinct \gls{fl} models, each tailored to a specific cybersecurity function. 
The first model is designed to detect potential vulnerabilities and exploits. 
It processes data encrypted using \gls{mhe}, ensuring that gradients shared among federation participants remain protected throughout the training process, presented in Figure~\ref{fig:mhe-dp-fl}. 
This encryption scheme guarantees that only authorized entities with the shared decryption key can access the global model, preserving data confidentiality.

The second model evaluates the system’s compliance with security patches and mitigation measures.
It uses threat intelligence data, audit logs, and forensic evidence to assess whether necessary defenses have been effectively implemented and maintained.

To enable secure and verifiable public sharing, the architecture incorporates a distributed ledger, with RepuCoin~\cite{yu2019repucoin} as the default implementation, though it can be substituted with alternative blockchain solutions as required. 
The ledger integration is abstracted via a well-defined interface, enabling substitution with alternative blockchain solutions as required, with minimal impact on the rest of the system.
During substitution, migration strategies such as dual-writing or secure data export/import are employed to ensure historical data integrity and compliance.

The distributed ledger records cryptographic hashes of \gls{cti} data, \glspl{zkp}, Encrypted Anonymized Threat Intelligence, smart contracts, and forensic audit logs. 
By leveraging blockchain technology, the system ensures data integrity, transparency, and tamper resistance, while privacy-preserving cryptographic mechanisms prevent unauthorized data exposure.

The architecture is designed to support knowledge sharing across three specific scenarios. 
In the first scenario, inter-organization sharing allows \gls{legion} systems across different organizations to collaborate, sharing threat intelligence and updating the local knowledge pool.
The second scenario considers the application of shared information from other organizations, combined with internal team data, each managing their own \gls{its}.
The third scenario enables public access, making relevant threat intelligence available to the broader cybersecurity community. 
This approach ensures open access to critical security data, fostering transparency, innovation, and enhanced threat mitigation strategies.\\

\noindent\textbf{Inter-Organization Sharing}: \gls{legion} facilitates federated collaboration at the organizational level, allowing different entities to securely exchange \gls{cis} and \gls{cti}. 
A primary challenge is maintaining privacy and confidentiality while ensuring that threat intelligence remains accessible to authorized participants within and beyond the federation.

\gls{legion} integrates \gls{misp} instances within each organization to address this, enabling structured, automated intelligence sharing. 
Privacy, confidentiality, and data integrity are preserved by encrypting \gls{cis}/\gls{cti} data using \gls{dp}, Encrypted Anonymized Data techniques, or \gls{mhe} before dissemination. 
These cryptographic mechanisms ensure organizations can securely exchange intelligence without exposing sensitive details.

Each organization enriches its intelligence repository by incorporating external intelligence from other federated entities alongside internally collected data from its local \gls{its} systems. 
By applying privacy-preserving techniques, shared intelligence remains protected even in the event of a compromise. 
The information cannot be linked to the original contributing entity nor exploited for further attacks, ensuring that threat intelligence remains secure, anonymized, and resistant to adversarial manipulation.\\

\noindent\textbf{Internal Sharing and Application}: Within an organization, exchanged threat intelligence, comprising both \gls{cti} and \gls{cis} data, is applied to each \gls{its} instance to support effective threat detection and ensure compliance with current security mitigation requirements.
Each \gls{its} independently generates intelligence based on local data and securely shares it with other \gls{its} instances across the organization.
Since individual \gls{its} deployments often serve distinct services, enforce different policies, and operate in varied environments, it is critical to prevent data leakage and to ensure that a compromise in one instance does not propagate to others.

To uphold security and privacy within the internal federation, all data exchanges adhere to the same cryptographic and privacy-preserving principles established for inter-organizational sharing.
Each \gls{its} processes only its telemetry and disseminates intelligence exclusively in encrypted or pre-structured formats designed for secure output transmission.
By leveraging encrypted gradients in combination with \gls{mhe}-based \gls{fl}, the system enables collaborative learning and information sharing without compromising confidentiality.

The resulting intelligence is applied across a variety of cybersecurity contexts.
\gls{mhe} is embedded into \gls{fl} workflows, supporting one model dedicated to threat detection, including vulnerability discovery, exploit detection, and identification of adversarial behavior.
A second model evaluates the effectiveness of implemented mitigation strategies using a combination of \gls{cti} inputs, encrypted audit logs, and forensic evidence.
\gls{zkp} further enhances these assessments by enabling the validation of security conditions, such as vulnerability exposure, without revealing sensitive underlying data.
Once generated, security assessment reports are transmitted to the \gls{soc}, where they inform actionable decisions for threat mitigation.

This process is inherently iterative, continuously refining insights based on newly collected data.
As illustrated in Figure~\ref{fig:arch}, several \gls{its} components contribute with critical telemetry, including Risk Managers, \gls{osint} databases, \glspl{hids}, firewalls, and security event logs from both \gls{vm} and hypervisor layers.
All collected data is subjected to privacy-preserving preprocessing before being securely transmitted to the distributed ledger or \gls{misp} and to the \gls{soc}.
The \gls{soc}, in turn, applies \gls{mhe}-enabled analytics to derive appropriate mitigation strategies, which are stored in the distributed ledger and \gls{misp} to support future detection, response, and resilience enhancement activities.\\

\noindent\textbf{Public Access}: The third and final scenario focuses on publicly disseminating collected threat intelligence data.
Integrating a distributed ledger within \gls{legion} enables external entities to utilize \gls{legion} as an additional \gls{osint} resource.

This approach has the potential to promote the widespread distribution of \gls{cis} and \gls{cti} data, enhancing situational awareness across the broader security community.
By providing timely insights into emerging threats and vulnerabilities, \gls{legion} supports proactive defense measures among interconnected systems. 
Additionally, assisting external entities in strengthening their security postures reduces the likelihood of \gls{legion} interacting with compromised or vulnerable environments. 
As a result, the network's overall resilience, security, and operational efficiency improve, contributing to a more robust cybersecurity ecosystem.

To prevent adversaries from adapting to disclosed intelligence, the distributed ledger in \gls{legion} stores only essential threat intelligence data while maintaining strict security and privacy controls. 
The integration of \gls{zkp} allows non-federated organizations to verify whether they are affected by vulnerabilities or exploits without revealing sensitive system details. 
This mechanism enables organizations to proactively seek mitigations and assess compliance with the latest security updates without exposing confidential information.

Smart contracts further enhance automation by triggering real-time alerts for participating organizations whenever newly identified threats require immediate action. 
This ensures rapid incident response and mitigation deployment.

The sharing of \gls{cti} data enables the security community to analyze potential indicators of malicious activity.
When combined with Encrypted Anonymized Threat Intelligence, these insights facilitate the identification of adversaries and the implementation of effective countermeasures.
Audit logs and forensic evidence also provide valuable metrics for threat hunting, augmenting \gls{ai} and \gls{ml}-driven automated detection methods to identify and respond to sophisticated cyber threats.

By leveraging the immutability and transparency of the distributed ledger while incorporating privacy-preserving cryptographic techniques, \gls{legion} ensures that adversaries cannot exploit shared intelligence.
Simultaneously, this strengthens the global cybersecurity community by providing verified, actionable intelligence without compromising sensitive organizational data, thus fostering trust, resilience, and long-term collaboration across the global cybersecurity landscape.

While these core components establish a strong foundation for federated threat intelligence, the evolving threat landscape demands further capabilities. 
To address emerging risks and operational complexity, \gls{legion} integrates additional specialized tools into its architecture.\\

\noindent\textbf{Additional Tools}: To further strengthen its capabilities, it is planned for \gls{legion} to incorporate a suite of advanced tools that extend beyond traditional threat intelligence collection and analysis. 
These tools are envisioned to be strategically integrated into key architectural components to provide comprehensive visibility, proactive defense, and automated response across the threat landscape.

\gls{legion} integrates an external attack surface management (ASM) module within the data collection and risk assessment layers, which continuously discovers and evaluates internet-facing assets, including misconfigurations, from an attacker’s perspective. 
By mapping the organization’s digital footprint in real time, ASM supports both the risk manager and the \gls{siem} pipeline to quickly detect and address newly exposed assets or vulnerabilities.

Digital risk protection (DRP) tools are incorporated into the external intelligence and monitoring layer, monitoring external sources such as the dark web and social media for mentions of the organization, leaked credentials, and brand impersonations. 
The DRP module feeds findings into the threat analysis workflows for early detection of attacks, breaches, or stolen information. 
This proactive intelligence, correlated with internal telemetry, enables timely mitigation actions and extends threat visibility beyond the enterprise perimeter.

To enrich investigations and threat detection, \gls{legion} supports the ingestion and correlation of global and local threat intelligence feeds. Platforms such as AlienVault OTX are integrated within the threat intelligence aggregation layer.
These feeds supply real-time indicators of compromise, malware signatures, and exploit activity contributed by a global community of researchers and organizations. 
The architecture ensures seamless integration of these feeds into the \gls{siem} and analytics pipelines, making the intelligence actionable and directly supporting automated incident response workflows.
The combination of global threat intelligence feeds ensures \gls{legion} benefits from collective knowledge, accelerating detection and response to new threats worldwide.

Advanced analytics, powered by \gls{ml} and \gls{ai}, are embedded in the analytics and decision-support layer. 
These technologies enable \gls{legion} to detect behavioral anomalies, predict emerging threats, and prioritize vulnerabilities based on real-world exploitation trends.
By continuously analyzing large datasets, \gls{legion} provides actionable insights and early warning alerts, enhancing situational awareness and supporting more informed decision-making.
\gls{ai} and \gls{ml} analytics help detect subtle anomalies and predict emerging threats.

Cyber risk management features are integrated into the risk assessment and governance components of the architecture. 
These tools facilitate ongoing evaluation and management of risks posed by supply chain partners, vendors, and third-party entities, extending \gls{legion}’s visibility and control beyond organizational boundaries. Considering the growing frequency and impact of supply chain attacks, this can be an important factor to be considered in the implementation.

Finally, \gls{legion} is designed to support automation and orchestration through integration with \gls{soar} tools.
Automation modules are embedded within the response and mitigation layer, enabling the system to execute predefined playbooks and automated responses to specific threats. 
This reduces detection-to-mitigation time and allows security teams to focus on higher-level strategic tasks. 
The architecture supports customizable intelligence models and automated workflows, ensuring that responses are tailored to the organization’s unique risk profile and operational requirements.

\subsection{Case Study: Federated Segmentation and Threat Intelligence Sharing}

To demonstrate the benefits of a federated, privacy-preserving framework for \glspl{its}, this subsection presents a case study illustrating how such an approach enhances threat mitigation, with a particular focus on segmentation and containment.

This case study examines CVE-2024-0132, a recently disclosed zero-day vulnerability in the NVIDIA Container Toolkit, which is widely used to enable GPU-accelerated containers in both cloud and on-premises environments. 
CVE-2024-0132 is caused by a time-of-check to time-of-use (TOCTOU) race condition, allowing a malicious container to bypass standard isolation mechanisms and gain unauthorized access to the host filesystem. 
Exploiting this flaw can result in arbitrary code execution, privilege escalation, data tampering, denial of service, and unauthorized disclosure of sensitive information.

The widespread adoption of the NVIDIA Container Toolkit across diverse platforms amplifies the potential impact of this vulnerability, threatening the heterogeneity and resilience that organizations and \glspl{its} often depend on.
While patches for CVE-2024-0132 have been released, unpatched or misconfigured systems remain at risk, highlighting the ongoing need for proactive detection and mitigation strategies.

Existing \glspl{its} often rely on intrusion detection systems such as Briareos~\cite{baptista2017briareos}, Suricata~\cite{fekolkin2015intrusion}, Bro IDS~\cite{sommer2003bro}, and Tripwire~\cite{kim1994design}.
However, their dependence on signature-based detection limits their effectiveness against zero-day threats, for which no known indicators or rules exist.
Complementary risk management frameworks, such as HAL 9000~\cite{freitas2024hal}, attempt to address this gap by recommending diversified configurations to reduce attack surfaces, promoting resilience through heterogeneity.
Nevertheless, these measures do not fully mitigate the risk of total compromise, especially in isolated or narrowly scoped \gls{its} deployments.

The federated architecture of \gls{legion} offers significant advantages in this context.
Its design enforces strict segregation between \glspl{its}, even within the same organization, limiting an adversary’s ability to move laterally after an initial compromise.
Rather than relying solely on direct detection and system-level defenses, \gls{legion} leverages controlled data flow, decoupling actionable intelligence from real-time access.
This approach gives \gls{siem} platforms critical time to evaluate anomalies and initiate responses before an attacker can escalate their activities.
Upon detecting a breach, the framework prioritizes the sanitization of shared intelligence to prevent contamination of the broader intelligence ecosystem.
This represents a deliberate trade-off: while traditional \glspl{its} prioritize speed and automated local response, \gls{legion} introduces an additional verification layer for shared \gls{cti}, ensuring that federation-wide collaboration remains trustworthy and resilient.

If a compromised \gls{its} contributes to the shared \gls{cti} corpus, it becomes essential to revoke and correct that data to preserve the trustworthiness and operational value of the entire federation. 
This underscores the need for revocation mechanisms to address compromised contributions.
While this research does not detail implementation specifics, we refer to studies on trusted \gls{cti} retraction and redactable blockchain-based intelligence systems~\cite{dunnett2022trusted, arikkat2025sectis, deuber2019redactable} as directions for operationalizing such capabilities in federated environments.

\gls{legion} aligns with the NIST Cybersecurity Framework 2.0~\cite{pascoe2023public}, reflecting a commitment to actionable intelligence and responsible governance.
The architecture automates key processes such as coordination, reporting, and data exchange, reducing human friction and accelerating response.
Automation is essential for maintaining resilience in operational environments where response time is critical for containment. 
\gls{legion} meets this requirement by enabling automated coordination and response workflows, while ensuring the integrity and privacy of shared intelligence throughout the federation.

These architectural decisions provide practical containment advantages under real-world threat conditions. 
In the context of CVE-2024-0132, \gls{legion} enables two clear containment outcomes.
For a single, possibly compromised \gls{its}, data from external sources remains protected by cryptographic safeguards, preventing attackers from exploiting or leaking sensitive threat intelligence. 
In scenarios involving multiple \glspl{its}, enforced segmentation, as depicted in Figure~\ref{fig:arch}, prevents cross-domain traversal, confining the attacker to the initially compromised domain and reducing the overall impact of the breach.

\subsection{Federated Learning: Non-\gls{dp} vs \gls{dp} Comparison}

To illustrate the practical considerations of implementing privacy-preserving techniques within the proposed architecture, this section presents an experimental evaluation of one of its core components. 
Specifically, we examine the integration of \gls{dp} into \gls{fl} to assess the resulting trade-offs between privacy and model utility.

The experiment aims to demonstrate that, in a federated environment where privacy and confidentiality are paramount, the exchange of model parameters among organizations or organizational units can be made resistant to re-identification and reverse-engineering attacks.

We use a subset of the anomaly-detection dataset from Han et al.~\cite{han2018anomaly}, which contains attack-free, fuzzy, and malfunction message records labeled as compromised (T) or non-compromised (R).
We compare two configurations: a standard \gls{fl} implementation and an \gls{fl} variant augmented with \gls{dp}.

The dataset is partitioned into 614,124 training samples and 546,333 test samples. 
Experiments run on a Debian 12 virtual machine under QEMU, provisioned with 62 GB RAM, a 32-core CPU, and an NVIDIA GeForce RTX 3090 GPU (24 GB VRAM). 
We extended the Flower framework~\cite{beutel2022flower}, adapting its PyTorch and Opacus examples to our dataset.

For the \gls{dp}-enhanced \gls{fl}, each client applies Opacus with a privacy budget of $\varepsilon=1.64$ and $\delta$=$10^{-5}$ ~\cite{abadi2016}.
All source code and processed data are available for access and visualization on our \href{https://github.com/tadeufrei/LegionPOC/}{GitHub page}.

\begin{table}[h!]
\centering
\begin{tabular}{|c|c|c|c|c|}
\hline
\textbf{Round} & \textbf{Setting} & \textbf{Accuracy} & \textbf{F1 Score} & \textbf{Recall} \\
\hline
\multirow{2}{*}{1} 
 & Non-\gls{dp} & 0.9737 & 0.9610 & 0.9651 \\
 & \gls{dp}     & 0.8211 & 0.7546 & 0.8253 \\
\hline
\multirow{2}{*}{2} 
 & Non-\gls{dp} & 0.9951 & 0.9934 & 0.9929 \\
 & \gls{dp}     & 0.8773 & 0.8279 & 0.8942 \\
\hline
\multirow{2}{*}{3} 
 & Non-\gls{dp} & 0.9837 & 0.9947 & 0.9948 \\
 & \gls{dp}     & 0.8811 & 0.8338 & 0.8955 \\
\hline
\end{tabular}
\caption{Federated Evaluation Metrics: Non-\gls{dp} vs \gls{dp}}
\label{tab:fed_eval}
\end{table}



\subsubsection*{Results interpretation}

Table~\ref{tab:fed_eval} presents the results comparing the standard \gls{fl} model and the \gls{dp}-enhanced \gls{fl} model. 
While the non-\gls{dp} model achieves higher accuracy and faster convergence, the \gls{dp} model provides formal privacy guarantees, reducing accuracy by 10-15\% across rounds.
Despite this trade-off, the DP model maintains strong predictive performance, with F1-score and recall steadily improving over time.



The privacy loss parameter $\varepsilon=1.64$ represents a moderate privacy guarantee within the differential privacy framework.
Lower values of $\varepsilon$, such as $\varepsilon<1$, correspond to stronger privacy protection, as they introduce more noise and make it more difficult for an adversary to infer the participation of any individual in the dataset. Conversely, higher values of $\varepsilon$, for example $\varepsilon>2$, are generally considered to offer weaker privacy guarantees and may increase the risk of information leakage~\cite{dwork2014,abadi2016,opacus2021}. 
At $\varepsilon=1.64$, a meaningful amount of noise is added to obscure individual contributions, mitigating risks such as record linkage or membership inference, while still allowing the model to retain sufficient accuracy for practical use.


Refining Opacus parameters, such as the noise multiplier and clipping norm, provides more control over the privacy-utility trade-off, critical for adapting privacy guarantees to varying use cases.

\section{Evaluation of \gls{legion}}\label{sec:evaluation}

To evaluate the functional requirements of \gls{legion}, we use the Scenario-Based Architecture Analysis Method (SAAM)~\cite{1019479, kazman2002scenario}.
This evaluation method examines how the architecture performs under expected operational conditions and adapts to potential future changes, highlighting its strengths and identifying areas for improvement in the context of federated threat intelligence.






\subsection{Integrating a New Internal Threat Intelligence Source}

Integrating a new internal threat intelligence source, such as an additional \gls{hids}, is a routine yet vital operation for organizations seeking to enhance their security posture.
The \gls{legion} architecture is intentionally structured with modular data collection and \gls{siem} integration layers, which significantly streamline the onboarding of new data sources.
When a new \gls{hids} is introduced, developers can create a dedicated connector or adapter that seamlessly interfaces with the existing data collection pipeline.
This ensures compatibility and smooth assimilation of telemetry data into the system.

Once the connector is in place, the incoming telemetry is subjected to established privacy-preserving mechanisms, such as data anonymization and pseudonymization, to ensure compliance with internal privacy policies and regulatory requirements. 
The data is then normalized and mapped to existing schemas, allowing it to be processed by analytics and threat detection workflows without disrupting established baselines or detection logic. 
This process may involve validating the new data format, updating parsing rules, and conducting integration testing to verify that the new source does not introduce inconsistencies or performance bottlenecks.

Operationally, the integration is managed through configuration adjustments and incremental rollout procedures, which allow for staged deployment and monitoring of the new data source’s impact on system performance and detection accuracy.
Automated monitoring tools track the health and throughput of the new connector, providing early warning of any anomalies or integration issues. 
If necessary, resource allocation and load balancing mechanisms are adjusted to accommodate increased data volumes, ensuring that the overall system remains responsive and reliable.

Importantly, this integration process does not necessitate changes to the core architecture, although it may require ensuring that the format of the new telemetry aligns with existing data standards and that the system can handle any additional processing load.
The modular and extensible design of \gls{legion} allows the organization to adapt to evolving internal security needs, enabling the rapid incorporation of new detection technologies and data sources without disrupting essential system functions or compromising operational stability.

\subsection{Sharing Anonymized Threat Intelligence with a Trusted External Partner}

Facilitating secure and privacy-preserving sharing of threat intelligence with external partners is a cornerstone of federated cybersecurity collaboration. 
\gls{legion} addresses this requirement through a two-level federation model that distinctly separates internal and external data sharing, allowing organizations to maintain granular control over what information is shared and with whom. 
The architecture leverages advanced cryptographic techniques, including \gls{mhe}, \gls{dp}, and \gls{zkp}, to ensure that any intelligence shared with external entities remains both anonymized and encrypted. 
These mechanisms collectively prevent the leakage of sensitive internal details while still enabling the dissemination of actionable indicators of compromise, attack patterns, and vulnerability information.

When onboarding a trusted partner, the process involves several coordinated steps. 
Intelligence is exchanged in a standardized format using \gls{misp} instances, which support structured, automated, and interoperable sharing of threat data. 
Privacy controls are configured to reflect the specific requirements and trust level of the partnership, such as limiting the granularity of shared data, applying additional anonymization to sensitive fields, or enforcing strict access controls. 
The architecture supports the negotiation and enforcement of sharing policies, allowing the organization to dynamically adjust what is shared based on evolving risk assessments or partnership agreements.

Operationally, the process is designed to be seamless and minimally disruptive. 
Integration with a new partner typically involves configuration adjustments within the existing \gls{misp} and privacy-preserving modules, without the need for modification of core system components. 
Ongoing monitoring and auditing capabilities are in place to track data flows and verify compliance with sharing policies, ensuring that the organization retains visibility and control over its shared intelligence. 
This approach allows \gls{legion} to support dynamic, policy-driven collaboration with a diverse range of external partners, while maintaining a strong security and privacy posture.

\subsection{Detecting and Responding to a Zero-Day Vulnerability Using Federated Learning}

The ability to rapidly detect and respond to zero-day vulnerabilities is critical in modern threat intelligence operations, where the window for effective mitigation is often extremely narrow.
\gls{legion} employs federated learning (\gls{fl}) models combined with privacy-preserving analytics to enable organizations to collaboratively train detection models without exposing sensitive raw data.
This approach allows each organization to contribute local threat intelligence, such as telemetry, behavioral indicators, and incident reports, while ensuring that proprietary or regulated information remains confidential.

In practice, each local \gls{its} instance processes its own telemetry and computes encrypted gradients or model updates, which are then securely aggregated into a global model. 
Advanced cryptographic mechanisms, including \gls{mhe}, are used to protect the confidentiality of model updates during transmission and aggregation. 
\gls{dp} techniques may be applied to further obscure individual contributions, mitigating the risk of inference attacks or data leakage.

The architecture’s analytics, risk management, and privacy layers are specifically designed to support such collaborative workflows.
Automated orchestration modules coordinate the training process, manage participant authentication, and enforce policy compliance. 
The system can dynamically adjust the federation, adding or removing participants as needed, based on organizational readiness, threat landscape changes, or resource constraints.

Operationally, the federated learning process is supported by robust monitoring and validation mechanisms, which track model convergence, detect anomalies, and ensure the integrity of the collaborative process. 
Once the global model is trained, it can be rapidly deployed across all participating organizations, enabling near real-time detection and mitigation of emerging zero-day threats. 
This scenario illustrates \gls{legion}’s capacity to support privacy-conscious, real-time collaboration and threat mitigation in highly dynamic, multi-organizational environments, while maintaining strict confidentiality and compliance with data protection requirements.

\subsection{Replacing the Default Distributed Ledger}

The architectural flexibility of \gls{legion} is evident when the system must replace its default distributed ledger with an alternative blockchain solution. 
The distributed ledger plays a central role in the architecture, recording cryptographic hashes of threat intelligence data, audit logs, and smart contracts to ensure data integrity, transparency, and tamper resistance. 
If the integration between \gls{legion} and the ledger is implemented through a well-defined abstraction layer (as described in Section~\ref{sec:architecture}), the transition to a new blockchain can be accomplished primarily by updating integration APIs and ensuring compatibility with the new protocols. 
This modular approach confines the impact of the change to the integration layer, minimizing disruption to the broader system.

However, if the interface is tightly coupled to the specific features or data structures of the original ledger, a broader set of modifications may be required across multiple components, potentially increasing the complexity and risk of migration. 
During the transition, historical data, including cryptographic hashes and audit logs, can be securely exported from the original ledger and re-imported into the new blockchain, with validation procedures in place to ensure completeness and integrity.

This scenario underscores the value of modular interfaces and architectural abstraction, which not only facilitate the adoption of new foundational technologies but also help localize the impact of such changes. 
Furthermore, careful planning and execution of migration strategies are crucial to preserving the integrity and accessibility of historical data, thereby ensuring ongoing compliance and auditability throughout the transition process.

\subsection{Extending the System to Support New Privacy-Preserving Technologies}

As cryptographic standards evolve, organizations may need to integrate new privacy-preserving technologies, such as post-quantum cryptography, into their security infrastructure. 
In the context of \gls{legion}, this process extends beyond simply swapping out cryptographic algorithms. 
Integrating fundamentally new primitives often necessitates updates to data exchange protocols, adjustments to key management schemes, and modifications to privacy-preserving workflows to accommodate the operational characteristics of the new cryptographic approach.

The modular design of \gls{legion}’s cryptographic layer is intended to minimize disruption during such upgrades.
This modularity allows for the isolation of cryptographic changes, enabling targeted updates without requiring a full system overhaul. However, the introduction of advanced cryptographic techniques, such as those designed to be resistant to quantum attacks, can introduce significant performance overhead due to larger key sizes and more computationally intensive operations.
This may affect system latency and throughput, especially in scenarios where real-time threat intelligence sharing is required.

Compatibility challenges may also arise, particularly in federated environments where multiple organizations must coordinate the adoption of new cryptographic standards. 
Ensuring interoperability across all federation members may require phased deployment strategies, temporary support for legacy algorithms, or the implementation of protocol negotiation mechanisms that allow systems to dynamically select appropriate cryptographic methods based on counterpart capabilities.

Operationally, integrating new cryptographic primitives requires careful validation to ensure that privacy and security guarantees are preserved. 
This often involves extensive testing of data flows, verification of protocol correctness, and reassessment of threat models to address any new risks introduced by the change. 
Additionally, updating documentation, retraining personnel, and revising compliance and audit procedures may be necessary to reflect the new cryptographic landscape.

By maintaining a flexible and modular cryptographic infrastructure, \gls{legion} is better positioned to incorporate emerging privacy-preserving technologies as they become necessary, while proactively managing the technical and organizational challenges that such transitions entail.

\subsection{Scaling to Support a Tenfold Increase in Participating Organizations}

Scaling \gls{legion} to support a tenfold increase in participating organizations, such as during a national or sector-wide deployment, introduces a range of architectural and operational challenges. 
The federation management layer must evolve to handle a significantly larger and more diverse set of participants, each potentially operating under different policies, network conditions, and regulatory constraints. 
This expansion places increased demands on the data aggregation pipelines, which must process and correlate a much higher volume of telemetry, alerts, and threat intelligence in near real-time. 
To prevent bottlenecks, data flows may need to be re-architected for greater parallelism, and aggregation points may require sharding or horizontal scaling.

The distributed ledger, which underpins data integrity and transparency, must also be enhanced to support higher transaction throughput and larger volumes of audit logs. 
This may involve optimizing consensus mechanisms, increasing block sizes, or deploying additional ledger nodes to maintain performance and fault tolerance. 
Federation management interfaces and orchestration tools must be improved to provide administrators with effective oversight, automated participant onboarding, dynamic policy enforcement, and granular monitoring across a much broader federation.

\gls{fl} models, which enable collaborative analytics and detection, may face scalability issues such as increased communication overhead, longer training times, and synchronization delays.
Addressing these challenges can involve adopting hierarchical or clustered federation structures, leveraging asynchronous or decentralized training protocols, and employing model compression techniques to reduce communication costs.

Proactive scalability planning is essential, encompassing the optimization of data storage solutions to accommodate larger datasets, the implementation of sophisticated load balancing strategies, and the development of robust federation management systems capable of supporting dynamic scaling. 
Continuous performance evaluation, stress testing, and architectural refinement will be required to ensure that \gls{legion} maintains high levels of performance, reliability, and security as it scales to meet the demands of a much larger and more complex federation.

\subsection{Handling a Large-Scale Security Incident}

In the event of a large-scale security incident, such as a coordinated cyberattack affecting multiple federated organizations, \gls{legion} must enable rapid and coordinated responses while upholding data integrity and privacy.
The architecture is designed to promote real-time threat intelligence feeds, allowing all impacted organizations to receive timely updates on \glspl{ioc}, attack vectors, and recommended mitigation actions.
Automated alerting and incident correlation mechanisms help identify the scope and progression of the attack, enabling security teams to prioritize and coordinate their response efforts.

Federation-based decision-making mechanisms are dynamically adjusted to facilitate information sharing and collective defense, while ensuring that sensitive data is protected through privacy-preserving analytics and cryptographic controls. 
The system supports the rapid dissemination of mitigation strategies, patch information, and forensic evidence, leveraging the distributed ledger for transparent and tamper-resistant event recording.

Operationally, \gls{legion} supports incident response playbooks and automated workflows that can be triggered in response to specific threat scenarios. 
These workflows may include the isolation of affected systems, the revocation of compromised credentials, and the deployment of emergency patches or configuration changes.
Throughout the incident, continuous monitoring and forensic logging are maintained to support post-incident analysis and compliance reporting.

This scenario tests the architecture’s ability to operate effectively under urgent, high-stakes conditions, requiring seamless coordination, robust automation, and swift action across the federation.
The system must balance the need for rapid response with the necessity to maintain privacy and data security, ensuring that collaborative defense does not come at the expense of organizational confidentiality or regulatory compliance.

\subsection{Integrating with External Threat Intelligence Platforms}

Integrating \gls{legion} with external threat intelligence platforms, such as commercial threat feeds, industry Information Sharing and Analysis Centers (ISACs), or public \gls{cti} repositories, can substantially enrich the system’s intelligence pool and enhance the detection of emerging threats.
The architecture includes a robust and extensible integration layer that supports the secure ingestion of external data streams, mapping and normalizing them to internal schemas for seamless analysis and correlation with locally collected telemetry.

To ensure operational security, all external feeds are subjected to rigorous validation, filtering, and sanitization processes before being incorporated into the threat intelligence workflow.
Privacy and data protection standards are strictly enforced, with sensitive or proprietary information either redacted or pseudonymized according to organizational policies and regulatory requirements. 
The system supports configurable trust levels for different external sources, enabling organizations to tailor their integration strategies based on the reliability and relevance of each feed.

Technical interoperability is achieved through the use of standardized data formats and APIs, such as STIX/TAXII, which facilitate automated, real-time exchange of threat indicators, signatures, and contextual information. 
The integration layer is designed to support dynamic onboarding and removal of external sources, allowing organizations to adapt quickly to changes in the threat landscape or to incorporate new intelligence providers as needed.

This capability highlights \gls{legion}’s extensibility and its readiness to interoperate with a diverse range of external intelligence providers, ensuring that organizations have access to the most current and comprehensive threat intelligence available for proactive defense and situational awareness.

\section{Discussion}\label{sec:discussion}

Integrating federated architectures into \glspl{its} introduces new opportunities and challenges in the context of collaborative threat intelligence and adaptive defense.
This section links the questions presented in Section~\ref{sec:introduction} regarding the deployment of federated paradigms in \gls{its} environments, drawing on insights from the design and operation of the \gls{legion} framework.

\subsection*{Q1: How can the federation paradigm be integrated into ITSs, and what challenges might arise?}

Federation can be integrated into \glspl{its} through a layered approach in which multiple independent \gls{its} instances, each responsible for specific services, departments, or organizational units, participate in a controlled, privacy-preserving information exchange mechanism. 
In \gls{legion}, this is achieved by enforcing segmentation and isolation policies at intra- and inter-organizational levels. 
Each \gls{its} maintains operational autonomy while participating in a broader collaborative environment where intelligence is shared selectively and securely.

Challenges arise in coordinating across heterogeneous systems, particularly in maintaining trust, ensuring data integrity, and reconciling differences in detection capabilities or reporting granularity.
Additionally, caution must be taken to prevent the propagation of false or compromised intelligence across the federation, which requires robust validation, revocation, and sanitization mechanisms, some of which remain areas of ongoing research.

\subsection*{Q2: How can information sharing be effectively managed at both inter-level (between entities) and intra-level (within an entity)?}

\gls{legion} manages inter-level information sharing through cryptographically protected communication channels, data sanitization pipelines, and the use of verifiable cryptographic constructs such as \glspl{zkp} and encrypted anonymized threat intelligence.
These techniques ensure that only vetted and de-identified intelligence is shared externally while preserving the privacy and policy constraints of the originating entity.

At the intra-level, \gls{its} instances within the same organization share intelligence through a secure internal federation. 
Local trust anchors facilitate this, and the same privacy-preserving mechanisms are applied at the inter-organizational level.
Segmentation policies enforced by the framework prevent lateral compromise while allowing \gls{fl} and cross-domain correlation to improve detection accuracy.
Information flow is governed by fine-grained access controls, ensuring that even within an entity, intelligence is shared only on a need-to-know basis.

\subsection*{Q3: What types of information can be safely and effectively shared between organizations without compromising privacy or security?}

\gls{legion} enables the sharing of a wide range of threat intelligence indicators while preserving confidentiality and source privacy.
This includes \glspl{ioc}, such as IP addresses, file hashes, malware signatures, and behavior patterns, as well as contextual metadata like timestamps, threat actor techniques, e.g., MITRE ATT\&CK mappings, and risk severity scores. 
Before dissemination, these artifacts are processed using techniques such as \gls{dp}, \gls{he}, and pseudonymization.

Using privacy-preserving mechanisms, sensitive fields such as host identifiers, user information, or internal configurations are redacted or obfuscated, preventing adversaries or other participating entities from inferring critical system details.
This balance between utility and privacy allows intelligence to retain its operational value without introducing new attack surfaces.

\subsection*{Q4: In what ways does newly acquired information benefit an individual entity within a collaborative system?}

New intelligence obtained through the federation enhances an individual entity’s situational awareness, improves detection capabilities, and facilitates more accurate threat attribution. 
This is especially valuable in the context of previously unseen or zero-day attacks, where collective insights can serve as early warning signals.
For example, disseminating early-stage \glspl{ioc} from a compromised organization may allow others to proactively harden defenses, update detection rules, or block malicious traffic before exploitation occurs.

Furthermore, intelligence derived from federated sources contributes to local risk assessment processes.
It allows entities to correlate external patterns with internal telemetry, enhancing threat hunting, forensic analysis, and decision-making around incident response.

\subsection*{Q5: How can shared information improve the overall resilience of a system or interconnected systems?}

System-wide resilience is significantly improved when intelligence is shared securely and responsibly across a federation. 
\gls{legion} facilitates this by enabling \glspl{its} to participate in distributed detection and defense workflows without exposing their internal state or compromising operational integrity.
Shared intelligence accelerates threat containment by enabling early detection and coordinated response efforts, reducing the time of mitigation across the federation.

By enabling automated but verified \gls{cti} exchange, \gls{legion} supports a defense-in-depth strategy where knowledge gained in one domain strengthens the defensive posture of others. 
This collective learning dynamic increases adversarial cost, reduces redundant analysis, and enhances long-term resilience by fostering a continuously evolving security baseline across interconnected systems.

\section{Conclusion}\label{sec:conclusion}

This research introduced \gls{legion}, a federated \gls{its} architecture designed to enhance cyber resilience through secure, privacy-preserving, and collaborative \gls{cti} sharing. 
By adopting the Federation paradigm, \gls{legion} enables structured knowledge exchange across intra-organizational and inter-organizational levels while preserving the autonomy and confidentiality of participating entities.

The architecture integrates distributed technologies, including distributed ledgers and \gls{misp}, to ensure the verifiability, traceability, and integrity of shared \gls{cti}.
It enforces fine-grained control over data dissemination through cryptographic safeguards such as \gls{mhe}, \gls{dp}, and \glspl{zkp}, enabling selective disclosure without exposing sensitive operational or contextual information.

Three operational scenarios were examined to illustrate the practical deployment of \gls{legion}: internal sharing within a single organization, cross-organizational collaboration, and controlled public dissemination of curated intelligence. 
In each case, the architecture supports the propagation of actionable knowledge while preventing adversaries from adapting to disclosed countermeasures or exploiting federation participants. 
These guarantees are maintained through enforced segmentation, structured revocation mechanisms, and controlled data flows.

The theoretical-practical scenario based on CVE-2024-0132 demonstrated how the framework enhances threat mitigation by decoupling real-time system response from intelligence-sharing workflows. 
This separation allows \glspl{siem} to verify and contextualize threat signals before federation-wide distribution, enabling organizations to respond effectively to zero-day threats without compromising data integrity or privacy. 
The approach further strengthens the deployment of secure and trustworthy \gls{ml}/\gls{fl} models by enabling them to learn from encrypted intelligence streams while remaining resilient to adversarial inputs.

In addition, the practical proof-of-concept integration of \gls{dp} into the \gls{fl} framework quantified the privacy–utility trade-offs inherent in these techniques.
The results demonstrate that the architecture can enforce privacy safeguards and effectively thwart re-identification attempts while maintaining competitive model accuracy and resilience.
These findings validate its practicality for deployment in real-world \glspl{its}.

\gls{legion} advances the state of the art by aligning operational efficiency with the principles of the NIST Cybersecurity Framework 2.0~\cite{pascoe2023public}, automating key workflows such as coordination, incident reporting, and \gls{cti} validation.
It provides a robust foundation for situational awareness and collaborative defense in environments where trust, privacy, and speed must coexist.

Future work will focus on implementing a functional proof-of-concept of \gls{legion}, validating its design through real-world experimentation, and evaluating its performance and scalability in federated, heterogeneous environments.
Further research will also explore adaptive trust models, redactable blockchain integration, and federated revocation policies to strengthen the reliability of collaborative intelligence ecosystems.

\section{Acknowledgements}
Tadeu Freitas was supported by Fundação para a Ciência e Tecnologia (FCT), 
Portugal (2021.04529.BD). 
Carlos Novo was supported by Fundação para a Ciência e Tecnologia (FCT), Portugal (2021.08532.BD).
This work is supported through the Operational Competitiveness and
Internationalization Programme (COMPETE 2030) [Project Atlas].

\section{Declaration of Generative AI and AI-assisted technologies in the writing process}

During the preparation of this work the author(s) used perplexity.ai in order to assist with drafting, improving clarity, and refining the language of the manuscript. 
After using this tool/service, the author(s) reviewed and edited the content as needed and take(s) full responsibility for the content of the publication.


\bibliography{blibliography}

\end{document}